\documentclass[pdflatex,sn-nature]{sn-jnl}
\usepackage{graphicx} 

\usepackage{multirow}%
\usepackage{amsmath,amssymb,amsfonts}%
\usepackage{amsthm}%
\usepackage{mathrsfs}%
\usepackage[title]{appendix}%
\usepackage{xcolor}%
\usepackage{textcomp}%
\usepackage{manyfoot}%
\usepackage{booktabs}%
\usepackage{algorithm}%
\usepackage{algorithmicx}%
\usepackage{algpseudocode}%
\usepackage{listings}%

\graphicspath{{Figures/}}

\begin{document}

\title{Designing sustainability under Earth-system rate constraints}

\author{Jos\'e Halloy$^{a,1}$, Petros Chatzimpiros$^{1}$, Fran\c{c}ois Graner$^{2}$, Thomas Gregor$^{3,4}$}

\affil[1]{Universit\'e Paris Cit\'e, CNRS, LIED UMR 8236, F-75013 Paris, France}
\affil[2]{Universit\'e Paris Cit\'e, CNRS, MSC UMR 7057, F-75013 Paris, France}
\affil[3]{Department of Stem Cell and Developmental Biology, CNRS UMR3738 Paris Cit\'e, Institut Pasteur, 25 rue du Docteur Roux, 75015 Paris, France}
\affil[4]{Joseph Henry Laboratory of Physics \& Lewis-Sigler Institute for Integrative Genomics, Princeton University, Princeton, NJ 08544, USA}

\abstract{Sustainability debates are monopolized by quantities: total energy use, material stocks, or recycling volumes. These metrics are informative, but they obscure a more fundamental constraint. On a materially closed but energetically open Earth, long-term viability is governed by rates: power density, areal power density, and the rate at which dispersed materials can be repurified under diffuse solar fluxes. We develop a physical framework that links technological metabolism, material dilution, and recycling limits through these rate constraints. This perspective clarifies why many contemporary technologies remain viable only through stock drawdown and high-power buffering, and why green growth strategies often shift rather than resolve the underlying mismatch between industrial activity and Earth-system dynamics. We argue that long-term sustainability requires both quantitative reduction in aggregate industrial throughput toward rate-bounded regimes and the development of technologies qualitatively compatible with biospheric operating conditions.}

\maketitle

\medskip
\medskip
\medskip
\medskip

\section{Sustainability as a physics-of-rates problem}

Contemporary sustainability debates overwhelmingly focus on quantities: total energy use, cumulative emissions, material stocks, or recycling volumes. While these metrics are informative, they often obscure a more fundamental constraint. On a materially closed yet energetically open Earth, long-term sustainability is a physics-of-rates problem, governed not only by how much matter and energy are used, but by the \emph{rates} at which they are transformed, dispersed, or repurified. Because the ecosphere is finite, rates of energy use, matter use, and waste production are necessarily coupled rather than independent.

\begin{figure}[t!]
  (A) \hfill (B) \hfill  ~\\
    \includegraphics[height=0.39\columnwidth]{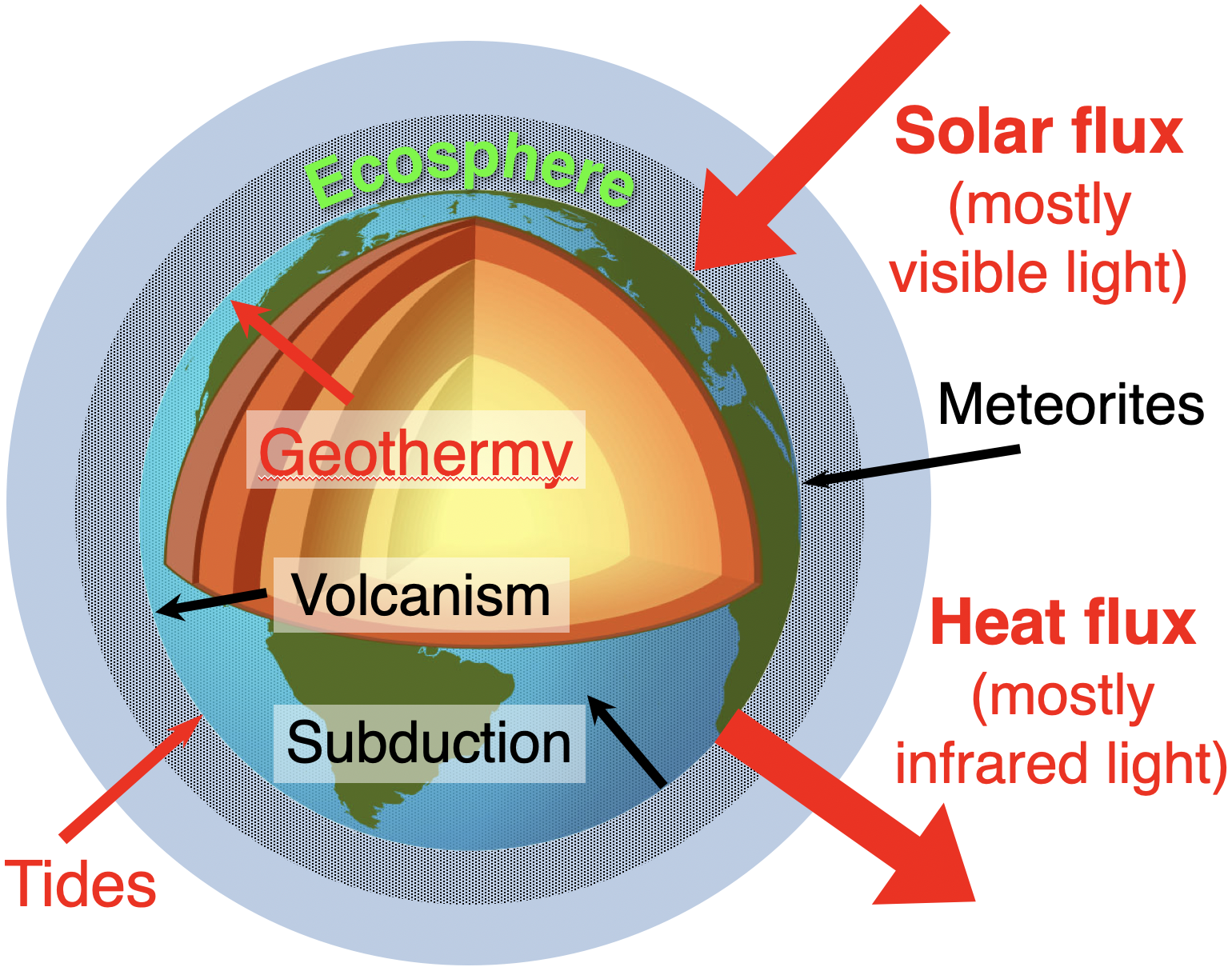} \hfill
    \includegraphics[height=0.39\columnwidth]{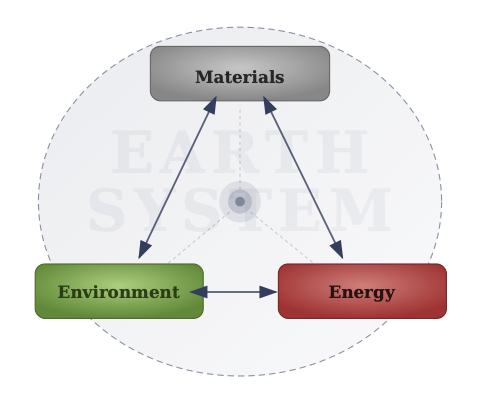} \hfill  ~
\caption{%
\textbf{Ecosphere finiteness couples energy, matter, and environment.}
\textbf{(A) Energy and matter fluxes in the ecosphere.}
The ecosphere, shown schematically and not to scale, occupies the thin interface of air, water, and soil that sustains life. Only net input and output fluxes are shown. Solar radiation dominates energy input, with minor contributions from geothermal heat and tidal friction; outgoing infrared radiation balances inputs at steady state. At relevant timescales, total ecosphere mass remains approximately stable, with net exchanges limited to processes such as meteorite infall, volcanic outgassing, and subduction. {\it Planet drawing credit: Wikimedia.}
\textbf{(B) Coupled constraints.}
Because the ecosphere is finite, changes in energy use, material transformation, or environmental state feed back on the others. Energy consumption, matter consumption, and waste production therefore cannot be treated as independent processes.
}
\label{fig:global_energy_and_matter_fluxes}
\end{figure}

\subsection{Earth system}

The importance of rates is evident in the functioning of the Earth system. Solar radiation reaches the top of the Earth's atmosphere as a diffuse and spatially distributed flux, with average power density of $\sim\!10^3~\mathrm{W\,m^{-2}}$ delivering a total $\sim\!10^{17}\,\mathrm{W}$. At the surface, maximum power density is typically below $\sim\!10^3~\mathrm{W\,m^{-2}}$, capping the areal energy flux available to sustain planetary processes~\cite{smil2017energy,Hermann2006}.

Matter is likewise constrained at planetary scale. The total mass of the ecosphere remains virtually stable. Volcanism and subduction generate large but internally balanced fluxes of order $\sim\!10^4~\mathrm{kg\,s^{-1}}$. Net exchanges with the exterior include meteorite infall, of order $\sim\!1~\mathrm{kg\,s^{-1}}$, whereas other fluxes---such as cosmic rays, solar wind, gas escape, or nuclear mass loss from natural radioactivity---are negligible at planetary scale.

Within these constraints, Earth sustains its major long-lived processes: climate homeostasis, biogeochemical cycling, and biological metabolism. Solar input, together with limited geothermal inputs and episodic high-energy events such as volcanism, drives water evaporation and the hydrological cycle ($\sim\!10^{16}$~W), atmospheric and oceanic circulation ($\sim\!10^{15}$~W), and global photosynthesis ($\sim\!10^{14}$~W)~\cite{smil2017energy,odum1971environment}. These processes are maintained not by concentrated energy sources, but by modest power densities distributed across vast areas and long timescales.

The dominant elements of life are carbon, hydrogen, nitrogen, oxygen, phosphorus, and sulfur (CHNOPS). Biological metabolism operates through large-scale biogeochemical cycles. Carbon, hydrogen, oxygen, and nitrogen circulate through atmosphere, hydrosphere, and soils at relatively low power density and over long timescales. Carbon cycles at $\sim\!10^7$~kg~s$^{-1}$, with volcanism contributing an additional $\sim\!10^4$~kg~s$^{-1}$~\cite{schramski2015human}. Nitrogen is fixed by biomass at $10^3$--$10^4$~kg~s$^{-1}$~\cite{Battye2017}. These cycles sustain biomass regeneration across a wide range of timescales. Phosphorus and sulfur also recycle, although more slowly, and are more tightly constrained by geological processes~\cite{andreae1990ocean,smil2000phosphorus,smil2013harvesting,schramski2015human}. Fossil accumulation, by contrast, occurred over geological timescales, producing long-term carbon storage in coal, oil, gas, and peat. 

\begin{figure}[t!]
  (A) \hfill  ~ \\
   \includegraphics[width=0.99\columnwidth]{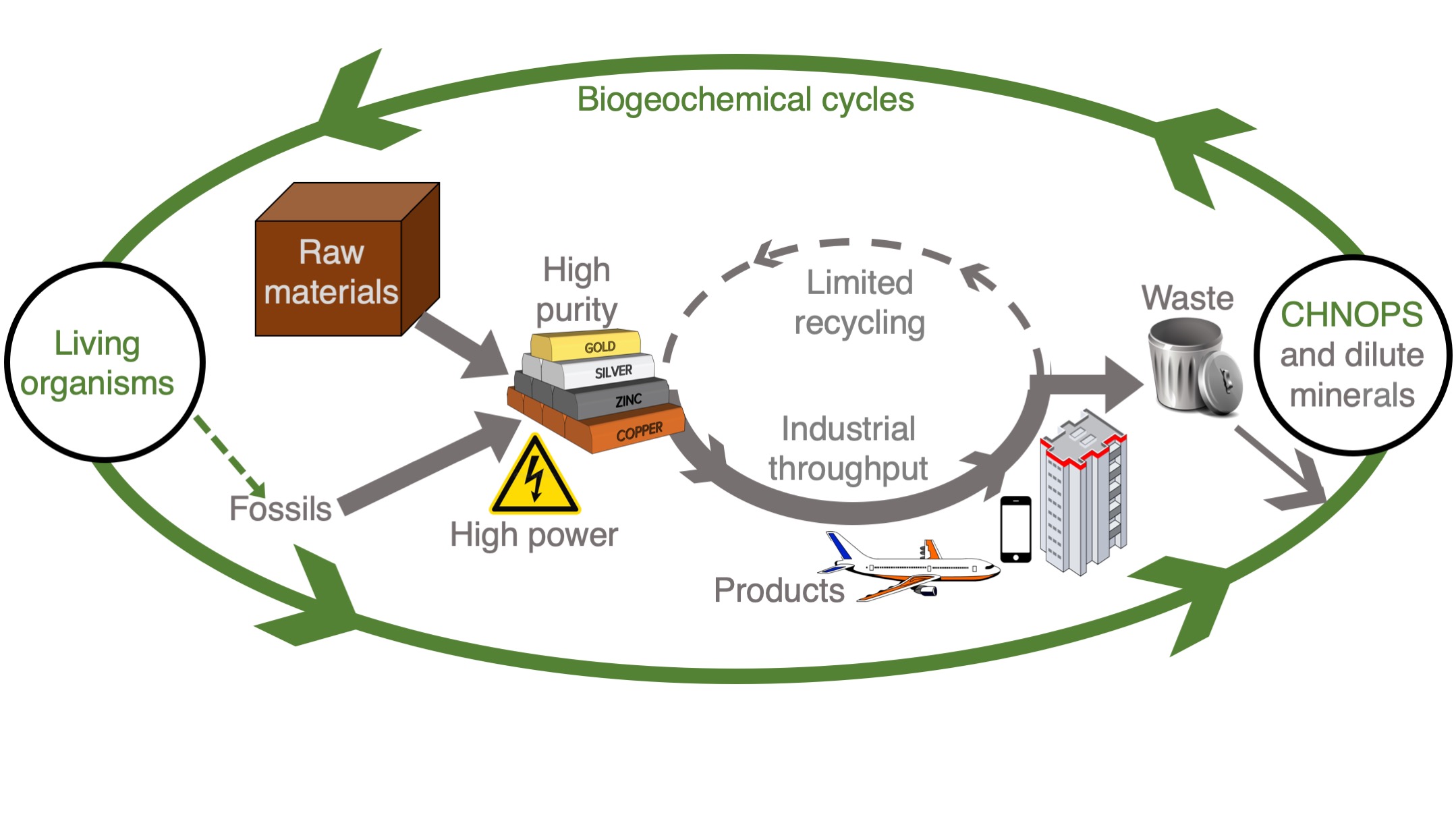}
   (B) \hfill  ~  \\
   \includegraphics[width=0.65\columnwidth]{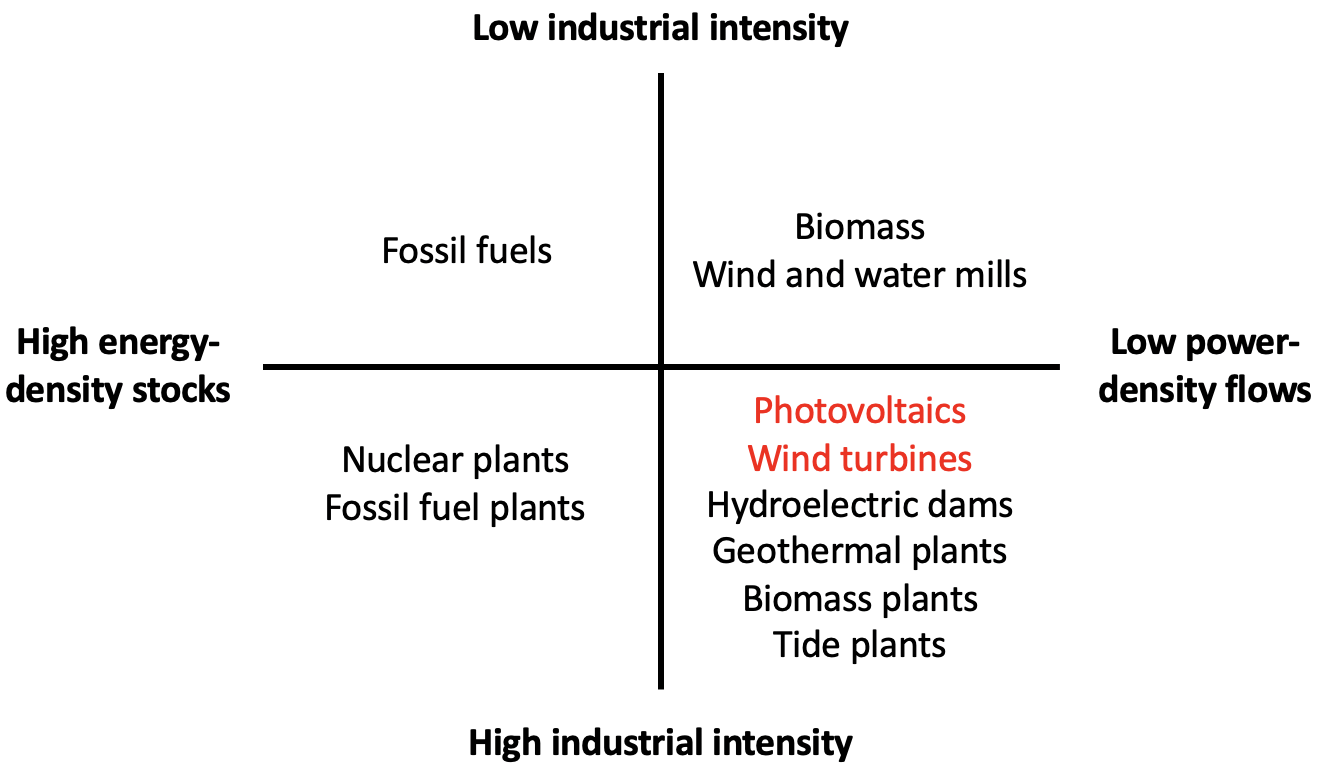} \hfill  ~
\caption{%
\textbf{Contrasting metabolisms of life and industry.}
\textbf{(A) Biological and technological metabolisms within the Earth system.}
Life and industry differ in pace, material logic, and coupling to planetary cycles.
\emph{Life metabolism:} solar flux drives biogeochemical cycles that recycle carbon (C), hydrogen (H), nitrogen (N), and oxygen (O), while phosphorus (P) and sulfur (S) are supplied from rocks at lower rates.
\emph{Technological metabolism:} industrial systems extract concentrated raw materials and transform them into high-purity ingots and complex devices through high-power processes. Their flows are largely linear, with limited recycling, rapid throughput, and substantial waste. Recycling chains are also bound by the same principles, involving multi-stage, high-purity and high-power processes.
\emph{Superimposed systems:} the biosphere and technosphere coexist within a materially closed but energetically open Earth, highlighting the mismatch between slow circular cycling and fast stock-buffered throughput.
\textbf{(B) Energy sources and industrial intensity.}
Energy systems and their corresponding converters are positioned according to their dependence on high-energy-density stocks versus low-power-density flows (horizontal axis) and their industrial intensity (vertical axis). Technologies highlighted in red, such as photovoltaics and wind turbines, illustrate that renewable energy conversion can still remain materially and industrially intensive.
}
\label{fig:metabolism}
\end{figure}

Figures~\ref{fig:global_energy_and_matter_fluxes} and~\ref{fig:metabolism}A summarize this planetary setting. The ecosphere is effectively a closed material system, with energy balanced between incoming solar and outgoing infrared radiation. Within this setting, the biosphere operates as a slow, largely circular metabolism powered by diffuse fluxes.

\subsection{Contemporary human society, energy and power}

Any human activity intended for long-term sustainability must ultimately align with this diffuse areal power-density ceiling, as well as with the limited rates at which dispersed materials can be re-concentrated.

Contemporary human society mobilizes total power of order $10^{13}\,\mathrm{W}$, which is a tiny amount compared to the aggregate solar influx. Instead, it is supplied through highly concentrated stocks --- typically fossil fuels --- and infrastructures that decouple power harvest from materials use and area~\cite{IEA2021,Hermann2006, Steinberger2010}. The distinction between \emph{energy} and \emph{power} is crucial. Energy quantifies capacity for action, whereas power determines \emph{what} can be done and \emph{how fast}. 
By building technologies on high power sources, industrial societies have rapidly generated high-throughput transformations of matter. With fossil fuels (followed by nuclear systems), energy stored over geological timescales was released in decades and on demand. This temporal compression enabled material transformations at rates far exceeding those sustained by the solar-driven Earth system~\cite{smil2017energy,Krausmann2018}.

The finiteness of stock resources such as fossil fuels and high-grade mineral ores is relevant in a specific way. Depletion of material resources~\cite{riondet2023applicability,prior2012resource,gordon2006metal} and fossil-fuel reserves~\cite{laherrere2022much} matters because declining ore grades and declining access to concentrated fuels reduce available \emph{power density} and increase the energetic and material overhead required for extraction, processing, and refining. In a rate-based framework, depletion acts both as an \emph{accelerant} and as a \emph{revealer}: it exposes the dependence of most industrial systems on concentrated stocks that temporarily mask their incompatibility with solar-limited operations ~\cite{hall2009eroi,Schramski2020,odum1971environment}.

From this perspective, many sustainability strategies fail not because they violate conservation laws, but because they violate \emph{rate constraints}. Recycling, substitution, and efficiency gains are not forbidden in principle; they are limited by the power, infrastructure, and organizational capacity required to perform repurification at scale and in time. Once this framing is adopted, climate change, biodiversity loss, resource depletion, and waste accumulation appear not as independent crises, but as coupled symptoms of a deeper mismatch between industrial activity and Earth-system metabolism~\cite{Lade2020}. These constraints couple molecular-scale chemistry, infrastructure-scale logistics, and planetary-scale power fluxes within a single, materially bounded system~\cite{Wolfram2026}.

\subsection{Technological metabolism and material mismatch}

To formalize this mismatch, it is useful to view intensive agricultural and industrial activity as a form of \emph{technological} or \emph{social} metabolism ~\cite{helga2015PNAS}. Like biological metabolism, it consists of coupled material and energy flows. Unlike biological metabolism, however, its structure, pace, and chemical logic are organized around purification, high concentration, and rapid throughput.

Concentrated energy sources enable the extraction of mineral resources, their transformation into high-purity materials, and their assembly into complex devices at rates that exceed the biospheric cycling by orders of magnitude ~\cite{Krausmann2018,elhacham2020global}. Figure~\ref{fig:metabolism}A contrasts these two metabolisms schematically and highlights their superposition within a single, materially closed Earth system. Industrial systems extract concentrated raw materials at roughly $\sim\!10^6$~kg~s$^{-1}$ and convert them into high-purity inputs and complex devices via high-power processes ($\sim\!10^{13}$~W). Their flows are largely linear, with incomplete recycling, high waste, and short turnover times.

The core problem is not only extraction, but also dilution. The rate limitation follows from practical thermodynamics rather than from abstract impossibility.
Industrial systems typically pass through a characteristic sequence: concentrated resources are extracted and refined into high-purity materials; these are then mixed and alloyed, miniaturized and integrated into more and more compact and high-complexity devices; finally they are dispersed across space, chemical forms, and use categories, ending up in infrastructures, products, and waste streams. 

The work required for purification scales with the entropy of mixing: the more diluted a material becomes, the more difficult it is to restore it to useful concentration~\cite{Georgescu1971}. Industrial repurification---the collection, sorting, separation, and refinement of materials into high-purity feedstocks---is therefore much more than the reverse of production. It requires not only energy in principle, but also logistics, institutional coordination, and sustained power to support collection, sorting, separation, and refining operations performed continuously and at scale. These operations are themselves constrained by infrastructure, labor, control systems, and material throughput. As dispersion increases, repurification must process larger volumes at lower concentrations, raising the required work per unit recovered material and slowing achievable rates. Repurification is therefore not impossible, but intrinsically rate-limited: beyond a certain throughput, dilution accumulates faster than recovery can proceed under renewable areal power density.

For a technical cycle to approximate biogeochemical closure, material recovery must be nearly complete at each turn of the loop. Once significant fractions of matter are dispersed, contaminated, or lost, continued reproduction of equivalent technical objects requires renewed inputs from primary extraction. This is why technological metabolism tends toward linearity even in systems nominally designed for circularity, including CHNOPS-based systems such as agriculture~\cite{Chatzimpiros2023}. When material flows exceed repurification capacity, recycling fractions saturate, stocks deplete, and residual waste accumulates. The mismatch is thus dynamic rather than conceptual: sustainability fails not because loops cannot be closed, but because they cannot be closed fast enough under diffuse energy fluxes~\cite{Wolfram2026,graedel2011recycling}.

The resulting metabolic mismatch is the coexistence of two fundamentally different regimes: a slow, circular, solar-powered biosphere and a fast, largely linear, stock-buffered technosphere. As long as industrial activity remains dominated by high-power-density processes and strategies built on high-purity feedstock materials, it will tend to disrupt biogeochemical cycles, deplete concentrated stocks, and accumulate dispersed waste faster than Earth-system processes can absorb or regenerate it.

\section{Zombie technologies and the limits of green growth}

Within this framework, many contemporary technologies can be understood as \emph{zombie technologies}: systems that remain operational and may even proliferate in the short to medium term, yet are physically incompatible with long-term viability. Their persistence is therefore not evidence of sustainability, but a transient consequence of stock drawdown, accumulated infrastructure, and geopolitical buffering~\cite{halloychp11}.

Zombie technologies share several defining features. They rely on finite energy or material stocks, demand sustained high power density, require continuous throughput to maintain functionality, generate complex and dispersed waste streams, and depend on repurification capacities that cannot be maintained under diffuse energy fluxes. Their apparent resilience thus masks an underlying metabolic insolvency.

The concept of green growth illustrates this tension clearly. In its strongest form, green growth assumes that technological activity can continue to expand while environmental impacts simultaneously decline~\cite{hickelisgreengrowthpossible,Parrique_EEB}. From a rate-based perspective, this expectation fails on two fronts. First, efficiency gains do not remove the thermodynamic degradation associated with dispersion and mixing as throughput increases. Second, large-scale substitution of fossil fuels by renewable energy systems typically requires expanded surface area, greater material intensity for energy capture, and additional infrastructure for storage and distribution. In that case, metabolic pressure is displaced rather than eliminated.

Figure~\ref{fig:metabolism}B illustrates that many ostensibly ``green'' technologies remain embedded in materially intensive, high-industrial-intensity systems even when they rely on renewable flows rather than fossil stocks~\cite{Schandl2024,graedel2015basis}. Electrification, digitization, and renewable deployment can reduce some emissions locally or temporarily, but they may also intensify demands on mineral extraction, land use, and repurification systems at global scale.

Zombie technologies are therefore not anomalies, but the predictable outcome of attempting to sustain high-throughput technological metabolism within a system constrained by diffuse energy fluxes and slow material cycles. 

\medskip
\textbf{A worked example: infrastructure turnover as a rate constraint.} Global production of cement and steel illustrates how infrastructure turnover functions as a rate constraint. These materials form the backbone of contemporary infrastructure~\cite{smil2022works}, and their manufacture requires high-temperature processes sustained by concentrated power sources. Globally, cement and steel production account for a large fraction of industrial energy use and material throughput~\cite{krausmann2017global,Schandl2024}. The relevant variable is not only the total stock of infrastructure, but also the rate at which it is expanded, demolished, and replaced. High turnover---driven by rapid urban expansion, short building lifetimes, and continuous reconstruction---amplifies extraction rates, energy demand, and waste generation. Even if individual processes become more efficient or partially decarbonized, accelerated replacement cycles increase aggregate throughput and the associated repurification burden. Conversely, extending infrastructure lifetimes through maintenance, adaptive reuse, and modular repair reduces the rate of material transformation. The distinction between zombie and viable regimes therefore does not hinge on abandoning concrete or steel as such, but on whether their production and replacement occur at rates compatible with long-term repurification under diffuse energy fluxes. The governing variable is thus not material identity alone, but the tempo of transformation. 
\medskip

This diagnosis is not the endpoint of the present Perspective. Its main purpose is to delimit the space of physically viable futures under Earth-system rate constraints. The central contribution of the following sections is therefore forward-looking: we propose a dual pathway combining quantitative reduction of aggregate throughput with the qualitative development of technologies compatible with biospheric operating conditions, and we examine the implications of this framework for transitions and global futures.

\section{A dual pathway: degrowth and life-analogous technologies}

We therefore propose a dual pathway toward viable futures: \emph{degrowth} of aggregate throughput toward a rate-bounded regime, and the development of \emph{life-analogous technologies} compatible with biospheric operating regimes. These two transformations are distinct yet inseparable. Only their joint implementation can align human activity with Earth-system constraints. 

Degrowth can encompass different economic, social, and cultural dimensions, but in all cases it encounters inescapable biophysical limits. Even total biomass is bounded by the availability of matter, especially CHNOPS, and by the conditions required to sustain its cycling. From the standpoint of a zombie metabolism, the relevant objective is therefore not endless expansion but stabilization at levels that prevent further zombification.

The longer-term objective is to move toward a technological metabolism that remains materially compatible with life. This calls for research-intensive efforts to phase out zombie metabolism and to invent, test, and scale technologies and metabolic loops better aligned with biospheric operating conditions.

\subsection{Degrowth as a correction of rates}

In the present framework, \emph{rate-bounded degrowth} denotes a decrease in the scale, speed, and power intensity of human activity sufficient to reduce aggregate material and energetic throughput into a range compatible with rate constraints~\cite{KALLIS2011873,dAlisa2014}. In a materially closed system, extraction, transformation, and waste generation must ultimately fall within the capacities of biogeochemical cycles or, alternatively, within engineered cycles that are themselves compatible with diffuse power and materially sustainable inputs. When industrial throughput exceeds these conditions, dilution outpaces recovery and the system becomes structurally unsustainable ~\cite{ONeill2018}.

In this framing, the Holocene can be understood as a bounded dynamical regime whose relative stability underwrote the development of human societies. The Anthropocene marks the disruption of that regime through technological throughput large enough to push Earth systems beyond the bounds that previously sustained that stability~\cite{steffen2007anthropocene,Rockstrom2009,steffen2015,steffen2018trajectories,Richardson2023,Rockstrom2024}.

\begin{figure}[t!]
    \centering
   (A) \hfill  (B) \hfill ~\\
   \includegraphics[width=0.49\columnwidth]{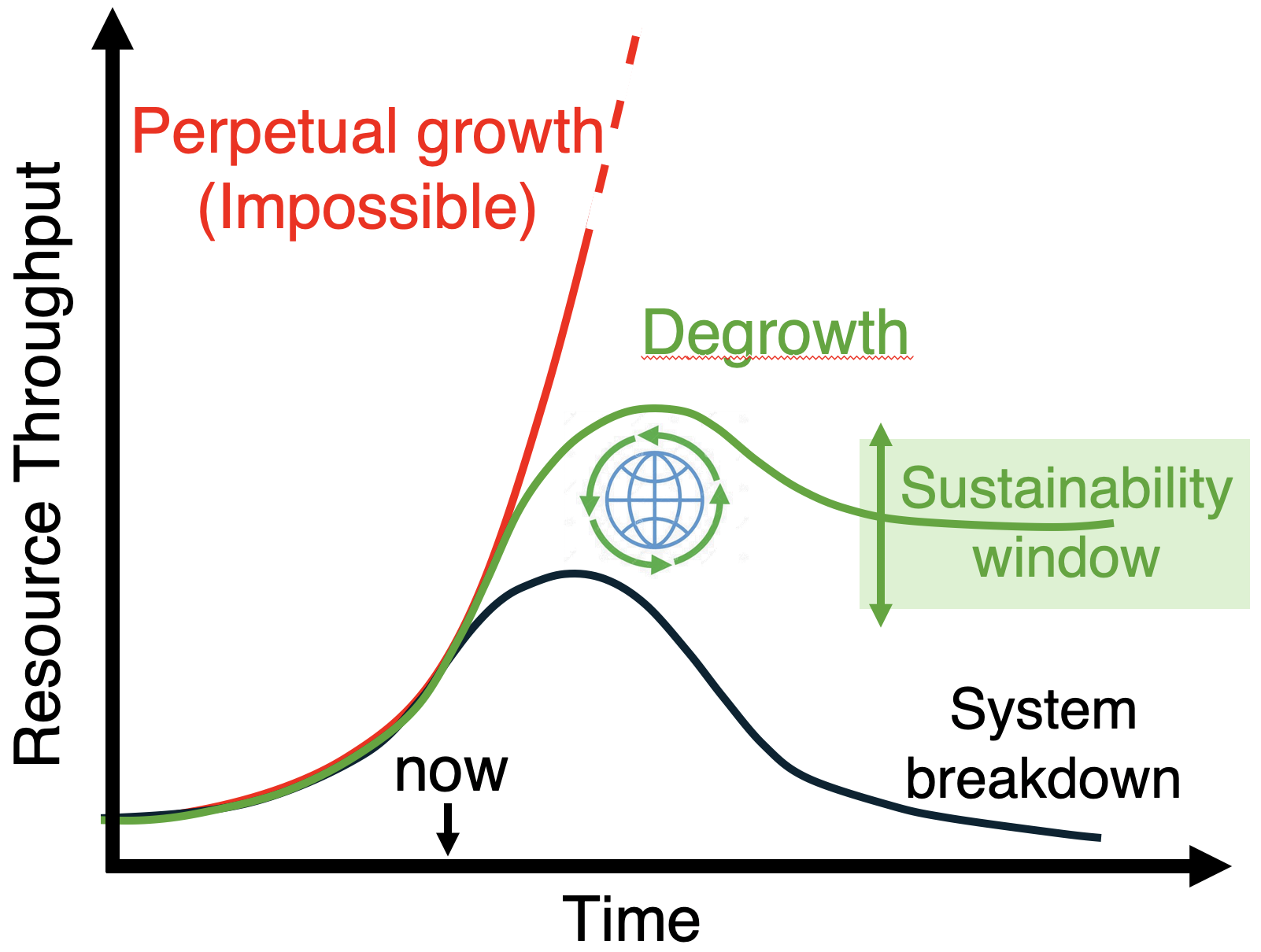}
   \includegraphics[width=0.49\columnwidth]{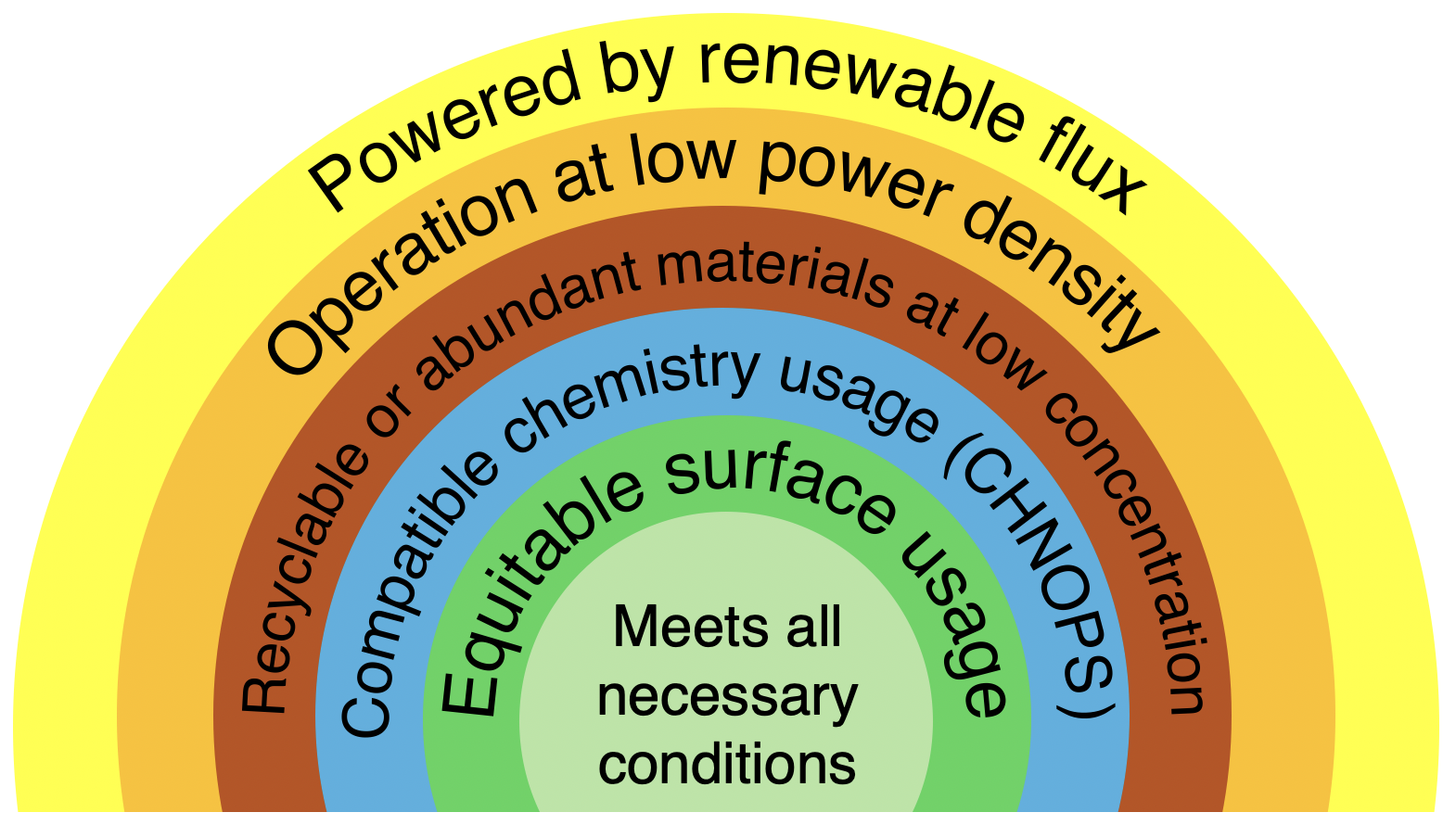}
\caption{%
\textbf{Degrowth and life-analogous technologies as dual requirements for sustainability.}
\textbf{(A) Degrowth as movement toward a bounded operating regime.}
Schematic trajectories for industrial society. Continued throughput growth (red) is physically unsustainable, whereas uncontrolled decline leads to collapse (black). A deliberate reduction in material and energy throughput (green) can move the system toward a bounded regime compatible with long-term biophysical constraints. Adapted from~\cite{meadows1972limits}.
\textbf{(B) Conditions for life-analogous technologies.}
Necessary conditions include renewable energy input, operation at low power density, use of recyclable or abundant materials at low concentration, compatibility with CHNOPS-centered chemistry where possible, and equitable surface use. Together, panels A and B summarize the dual pathway proposed here: reducing throughput while redesigning technologies to operate within Earth-system rate constraints.
}
\label{fig:dual_pathways}
\end{figure}

From this perspective, rate-bounded degrowth is a \emph{transient process} through which aggregate power use and material throughput are deliberately reduced until they enter a bounded regime compatible with long-term maintenance under solar-limited power densities. Figure~\ref{fig:dual_pathways}A schematizes this logic. Continued exponential growth in throughput is physically impossible over long timescales, while uncontrolled decline leads to breakdown. Between these extremes lies a family of \emph{rate-bounded operating regimes}~\cite{daly1977steady} in which fluctuations remain bounded rather than divergent. The exact boundaries of this regime remain uncertain; determining and quantifying them is an important and largely unexplored sustainability problem.

Crucially, the required reduction is \emph{absolute}, not merely relative. Improvements in efficiency, decarbonization intensity, or recycling fractions are insufficient if total throughput continues to rise. Because matter and energy degrade in quality through dispersion and mixing, reductions must be large enough that repurification can proceed at rates compatible with available power per unit area. This implies reductions not only in total energy use, but also in the \emph{speed, scale, and concentration} of industrial activities: extraction, transport, manufacturing, and replacement.

Seen in this way, rate-bounded degrowth is not primarily a moral or economic claim, but a physical one. Material fluxes must slow to timescales that allow biological, geochemical and infrastructural feedbacks to operate effectively~\cite{Wolfram2026}. High-power industrial processes---especially those relying on extreme purity, rapid turnover, and continuous expansion---cannot remain viable in the long run under a diffuse areal power-density ceiling. Their contraction arises from the physics. By contrast, activities centered on longevity, reuse, maintenance, and repair become central because they reduce throughput without requiring corresponding increases in repurification intensity.

This emphasis on bounded regimes should not be confused with a static vision. Complex systems can remain dynamic and adaptive while their fluctuations remain bounded. The framework therefore excludes trajectories based on indefinite escalation of power density and material throughput, without prescribing a single social or institutional outcome.

Degrowth alone, however, is not sufficient. Quantitatively reducing throughput without qualitatively transforming the material and technological logic of the system would merely shrink an incompatible metabolism without resolving its structural constraints. This leads directly to the second part of the dual pathway.

\subsection{Life-analogous technologies}

Long-term sustainability should rely on technologies capable of operating indefinitely within constraints comparable to those that have governed the biosphere for billions of years. We refer to such systems as \emph{life-analogous technologies}~\cite{halloychp11}. The biosphere shows that persistent, complex organization is possible under diffuse energy fluxes, low power density, and slow material cycling~\cite{vernadsky1998biosphere}. While life is not perfectly circular and has transformed Earth irreversibly over geological time, its persistence reflects a deep compatibility with Earth-system rates.

Life-analogous technologies are defined by their fit within this \emph{operational envelope}. The central criterion is not novelty but \emph{rate alignment}: extraction, operation, maintenance, and end-of-life reintegration must all remain compatible with power and material fluxes sustained by diffuse solar input. Figure~\ref{fig:dual_pathways}B summarizes necessary, though not always sufficient, conditions.

First, they must be powered predominantly by renewable energy fluxes rather than by finite energy stocks. Second, they must operate at low power density and low areal power density, compatible with the per-square-meter ceiling imposed by solar irradiance. Technologies that require sustained high power density tend to depend either on concentrated fuels or on extensive infrastructures whose construction and maintenance reintroduce metabolic mismatch.

Third, life-analogous technologies should rely primarily on abundant or genuinely recyclable materials, with metals used sparingly and, where possible, in readily separable forms. Industrial dependence on highly purified, multi-element alloys and complex composites sharply increases repurification burdens at end of life. By contrast, living systems rely overwhelmingly on a small set of abundant elements circulating through interconnected multi-scale cycles~\cite{smil2013harvesting,schramski2015human}; non-abundant elements are used in dilute form, while enzymatic catalysis achieves high efficiency without requiring costly reconcentration stages. Technologies whose material strategies remain closer to such cycling logics, or whose components can be readily disassembled and reused, place lower demands on repurification rates.

Fourth, land and surface use must be considered explicitly. Harvesting diffuse renewable power requires area, and scaling technologies by expanding surface footprints can generate new conflicts with ecosystems, food production, and other human uses.

Some existing technologies already satisfy part of this logic. Passive solar heating, solar thermal systems, small-scale wind- or water-powered devices, earthen architecture, and low-energy food preservation practices all operate at relatively low power density and with modest material requirements. Even these systems, however, can become incompatible if scaled without regard to material sourcing, land use, or maintenance capacity.

Looking forward, several research directions point toward richer possibilities. These include self-repairing materials, systems designed to tolerate slowness and variability, and infrastructures that integrate biological processes more directly. Mimicking or harnessing biological processes is one route toward life-analogous technologies, but not the only one~\cite{halloy2018sustainability,halloychp11}. 

Life-analogous technologies alone, however, are not sufficient. Qualitatively transforming the material and technological logic of the system without quantitatively reducing absolute global throughput would still leave the metabolism unsustainable over the long term. Life-analogous technologies therefore need to be coupled to degrowth.

\section{Implications for global futures}

Rate-bounded degrowth and life-analogous technologies are mutually reinforcing. Throughput reduction moves industrial activity toward a rate-bounded regime, while life-analogous technologies displace zombie dynamics. Together, they define a constrained but non-trivial design space for long-term sustainability.

\subsection{Transitions as metabolic regime shifts}

Most transition scenarios implicitly assume continuity: existing levels of service are preserved while energy sources and technologies are replaced. From a rate-based perspective, this assumption is fragile. Fossil-fuel systems concentrate power in space and time, enabling high-throughput technological metabolism through relatively compact infrastructures. Sustainable energy systems, by contrast, rely on diffuse fluxes and therefore generally require larger spatial footprints and broader material envelopes to deliver comparable services. Transitions that aim to preserve aggregate throughput may therefore reproduce metabolic mismatch even under a new energy regime.

Recognizing transitions as metabolic regime shifts highlights the importance of tempo and sequencing. Rapid expansion of renewable capacity without parallel reductions in material throughput can temporarily reduce some emissions while intensifying pressures on land use, mineral extraction, and recycling systems. The challenge is not to identify a single optimal pathway, but to navigate a constrained design space in which rates of deployment, contraction, maintenance, and repurification remain mutually consistent.

\subsection{Circularity under rate constraints}

Circular economy strategies occupy a central place in sustainability discourse~\cite{Stahel2016,Desing_circular_economy}, yet they are often evaluated through mass balances or recycling fractions rather than through power and throughput requirements. From a rate-based perspective, circularity is not a binary property but a bounded performance determined by the energy, logistics, and organizational capacity required for collection, sorting, separation, and reprocessing.

This has two implications. First, recycling cannot compensate indefinitely for high-throughput, high-dispersion systems; beyond a certain scale, repurification lags behind dilution~\cite{graedel2011recycling,reck2012challenges}. Second, circularity is often improved more effectively by slowing flows---extending product lifetimes, reducing complexity, and standardizing components---than by intensifying recycling itself. Policies and metrics that celebrate recycling fractions while ignoring absolute throughput risk masking rate mismatches rather than resolving them.

\subsection{Research and design frontiers under rate constraints}

In growth-oriented paradigms, innovation is often tied to higher performance, miniaturization, and increased functionality per unit cost. Under Earth-system constraints, the relevant question becomes whether technologies remain compatible with low power density, long timescales, and maintenance-oriented operation.

This shift elevates properties that are often undervalued in conventional metrics and value chains: repairability, robustness, ease of disassembly, tolerance to variability, and institutional capacity for long-term stewardship. It redirects design away from maximizing throughput and toward maintaining complexity under diffuse energy fluxes without accumulating irreversible material debts.

This perspective opens several research frontiers. One concerns measurement: current indicators emphasize efficiency or recycling fractions, but few assess whether material and energetic flows are compatible with repurification rates under diffuse power-density constraints. Developing metrics of \emph{rate alignment} between throughput, maintenance, and regeneration remains an open challenge.

A second frontier concerns technological design. Life-analogous technologies suggest criteria distinct from conventional optimization: tolerance to low and intermittent power density, reduced chemical complexity, modularity, and long service lifetimes. This shifts emphasis from peak performance towards robustness, repairability, and graceful degradation, while raising new questions about scalability and limits.

A third frontier concerns modeling. Rate-based approaches call for models that track the coupled dynamics of power density, material dispersion, repurification capacity, and land use across scales. Together, these frontiers define a scientific agenda centered on regime compatibility rather than marginal optimization.

\subsection{Avoiding re-zombification}

The framework developed here also highlights a central risk for future transitions: \emph{re-zombification}. Technologies initially compatible with diffuse renewable fluxes can become zombie-like if they are scaled in ways that demand escalating material throughput, concentrated power, or continuous replacement. Digital infrastructures, large-scale renewable installations, and even bio-based systems can fall into this trap if rate constraints are not explicitly considered.

Avoiding re-zombification requires embedding power-density and throughput criteria directly into design, assessment, and governance. Technologies should be evaluated not only by operational emissions or local efficiencies, but by their full metabolic signature: extraction rates, spatial footprint, maintenance demands, and end-of-life reintegration. Matter, energy, and waste flows must be considered in absolute planetary terms, not only per capita or per device. Otherwise, transitions may yield short-term gains while locking societies into new forms of long-term unsustainability.

\section{Conclusion: designing under an areal power-density ceiling}

Sustainability on a finite planet is ultimately a problem of design under constraint. The Earth system imposes a per-square-meter ceiling on usable power and a limited capacity to repurify dispersed materials within relevant timescales. Industrial systems that ignore these constraints can persist only temporarily, by drawing down accumulated stocks and externalizing metabolic costs. Long-term viability requires aligning human activity with rates the planet sustains.

The dual pathway outlined here---quantitative throughput degrowth toward a rate-bounded regime and qualitative development of life-analogous technologies---provides a coherent framework for such alignment. Together, these two requirements reframe sustainability not as maintaining existing industrial dynamics with cleaner inputs but as rebuilding technological metabolism within a bounded operating envelope.

Institutional and infrastructural dimensions are central to this reorganization. Maintenance capacity, repair skills, and logistical coherence are critical to long-term viability, yet are rarely treated as infrastructure in their own right. The biosphere demonstrates that persistent complexity under diffuse energy fluxes is possible. Its persistence does not derive from growth maximization, but from compatibility with planetary rates.

Translating that lesson into technological and social systems is a formidable challenge, but it defines a clear research and design frontier: identifying, testing, and institutionalizing ways of living and producing that remain viable under an areal power-density ceiling and finite material-cycling capacities.

\medskip
\medskip

{\it For the purpose of Open Access, a CC-BY 4.0 public copyright licence
\url{<https://creativecommons.org/licenses/by/4.0>} has been applied by the authors to the present document and will be applied to all subsequent versions up to the Author Accepted Manuscript arising from this submission.}

\bibliography{sustainability}

\end{document}